\documentclass[pre,twocolumn,superscriptaddress,showpacs,amsmath,amssymb]{revtex4}

\usepackage{graphicx}
\usepackage{bm}

\begin{document}

\bibliographystyle{apsrev}

\title{Effective triplet interactions in nematic colloids}

\author{M. Tasinkevych}
\affiliation{Max-Planck-Institut f\"{u}r Metallforschung,
Heisenbergstrasse 3, 70569 Stuttgart, Germany} \affiliation{Institut
f\"{u}r Theoretische und Angewandte Physik, Universit\"{a}t
Stuttgart, Pfaffenwaldring 57, 70569 Stuttgart, Germany}

\author{D. Andrienko}
\affiliation{Max-Planck-Institut f\"{u}r Polymerforschung,
Ackermannweg 10, 55128 Mainz, Germany }

\begin{abstract}
Three-body effective interactions emerging  between  parallel
cylindrical rods immersed  in a nematic liquid crystals are
calculated within the Landau-de~Gennes free energy description.
Collinear, equilateral and midplane configurations of the three
colloidal particles are considered. In the last two cases the
effective triplet interaction is of the same magnitude and range
as the pair one.
\end{abstract}

\date{\today}

\maketitle

\section{Introduction}
In colloidal suspensions knowledge of particle interactions is the
key to understanding their thermodynamic properties. Typically it
is assumed that the main contribution comes from the pairwise
interactions between the particles. This simplifies the
corresponding equations of motion which can then be solved using
efficient numerical algorithms, such as molecular dynamics.
However, in many systems the governing physical equations are
nonlinear and this, in principle, can result in a modification of
the interaction between two particles by the presence of other
ones, meaning that the total interaction energy is no longer
pairwise additive, and additional many-body potentials appear. The
importance of many-body interactions has been recognized in a
number of systems, from electron screening in
metals~\cite{hafner87}, catalysis and island formation on
surfaces~\cite{osterlund99}, to charged
colloids~\cite{russ02,brunner04,russ05,dijkstra03,hynninen03},
binary mixtures~\cite{amokrane05}, colloidal
crystals~\cite{dobnikar03,dobnikar03a} and simple
liquids~\cite{russ03,zahn03}.

Attempts to characterize the solvent-mediated interactions have
been made in case of nematic
colloids~\cite{ruhwandl97.0,lubensky98,andrienko:2003.a}, where
colloidal particles experience long-ranged interactions mediated
by the distortions of the nematic director around the
particles~\cite{ramaswamy96,poulin97,yada04}. These effective
interactions result in particle clustering and
self-organization~\cite{poulin97,anderson01}. Due to the presence
of topological
defects~\cite{terentjev95,kuksenok96,ruhwandl97.1,stark99,andrienko:2001.b}
the interparticle interaction is inherently anisotropic: depending
on the symmetry of the director field around the particles either
dipole-dipole or quadrupolar-type interactions can be
observed~\cite{poulin97,loudet01}.

The long-ranged nature of interactions mediated by the nematic
director can result in a substantial many-body contribution to the
interaction energy already
 for small concentrations of colloids. Indeed, recent theoretical studies of $N$-mers~\cite{gupta05} and trimers~\cite{guzman04} dispersed in a nematic
 host have shown that the three-body interactions can be of the same importance as two-body ones.

In spite of this fact, most research on nematic colloids still
relies on pairwise interaction potentials between the
particles~\cite{lev99,fukuda02,chernyshuk05,andrienko:2005.a,andrienko:2004.a}.
Moreover, often the director distribution around $N$ particles is
approximated by a superposition of one-particle solutions of a
corresponding linearized theory (so-called Nicolson
approximation), which yields vanishing many-body potentials.
In addition, a linear superposition of one-particle solutions does not satisfy boundary conditions at all
particle surfaces.

In this paper we would like to address the role of three-body
interactions for long cylinders immersed in a nematic mesophase.
This system can serve as a two-dimensional representation for
nanoparticles in a nematic host. It is also a realistic model for
carbon nanotubes-nematic liquid crystal composites (discotic
carbonaceous mesophases), which have recently been studied
theoretically~\cite{gupta05} and experimentally~\cite{zimmer83}.

\section{Many-body potentials and nematic free energy}
\label{sec:free_energy}

Due to the anisotropic nature of the solvent, an analytical treatment of particle interactions is practically impossible already at a pairwise level. We therefore choose here a numerical approach: equations for the nematic order tensor are solved by minimizing the free energy numerically using finite elements with adaptive meshes \cite{zienkiewicz05}. For a system of $N$ colloids at  positions ${\bm r}_i$ we introduce an excess free energy $F_N$ (defined below) over the free energy of a uniaxial nematic. $F_N$ is  then decomposed into effective $n$-body potentials $F^{(n)}$, with $n\leq N$, as following~\cite{russ02}
\begin{equation}
F_N = NF^{(1)} + \sum_{i<j}^{N}F^{(2)}(i,j) + \sum_{i<j<k}^NF^{(3)}(i,j,k)+ ...,
\label{eq:el_free_en_decompos}
\end{equation}
where we introduce the short-hand notation for the center of mass
positions $({\bm r}_i,...) \equiv (i,...)$.  The $n$-body
potential is defined in the system with $n$ colloidal particles.
$F^{(1)}= F_1$ is the excess free-energy coming from one isolated
colloid. The effective pair interaction between two colloids
follows from Eq.~(\ref{eq:el_free_en_decompos}) when $N=2$
\begin{equation}
F^{(2)}(1,2) = F_2(1,2) - 2F_1.
\label{eq:pair}
\end{equation}
The triplet potential $F^{(3)}(1,2,3)$ is defined in the
system of three colloids, $N=3$
\begin{equation}
F^{(3)}(1,2,3)=F_3(1,2,3) - \sum_{i<j}^{3} F^{(2)}(i,j) - 3F_1.
\label{eq:ternary}
\end{equation}
By construction $F^{(3)}(1,2,3)$ tends to zero whenever one of the
$r_{ij}\equiv \vert{\bm r}_i - {\bm r}_j\vert\rightarrow\infty$.
$F^{(3)}(1,2,3)$ is the quantity
 we are interested here.

The nematic host is modelled by the Landau-de Gennes free energy
density~\cite{degennes95}
\begin{eqnarray}
\nonumber
f &=& a(T-T^*)Q_{ij}Q_{ji} - bQ_{ij}Q_{jk}Q_{ki} + c(Q_{ij}Q_{ji})^2 \\
&+& \frac{L_1}{2} Q_{ij,k}Q_{ij,k} + \frac{L_2}{2}Q_{ij,j}Q_{ik,k},
\label{eq:nem_free_en}
\end{eqnarray}
where $Q_{ij}$ is the order parameter tensor, summation over
repeated indices is implied and the comma indicates the spatial
derivative. The positive constants $a,b,c$ are assumed to be
temperature independent, and $T^*$ is the supercooling temperature
of the isotropic phase.  The constants $L_1$ and $L_2$ are related
to the Frank-Oseen elastic constants $K_{11} = K_{33} = 9Q_b^2(L_1
+ L_2/2)/2$, $K_{22} = 9Q_b^2L_1/2$, and $Q_b$ is the bulk nematic
order parameter. We introduce the dimensionless temperature $\tau
= 24ca(T-T^*)/b^2$. The bulk nematic phase is stable for $\tau <
\tau_{\rm NI} = 1$ with a degree of orientational order given by
$Q_b = b(1+\sqrt{1-8\tau/9})/8c$; $Q_b(\tau>1)=0$.

For the constants entering the free energy density
(\ref{eq:nem_free_en}) we use typical values for a nematic
compound 5CB ~\cite{coles.hj:1978,kralj.s:1991}: $a = 0.044 \times
10^6 \,\mathrm{J/m^3K}$, $b = 0.816 \times 10^6 \,\mathrm{J/m^3}$,
$c = 0.45 \times 10^6 \,\mathrm{J/m^3}$, $L_1 = 6 \times 10^{-12}
\,\mathrm{J/m}$, $L_2 = 12 \times 10^{-12} \,\mathrm{J/m}$, $T^* =
307 \,\mathrm{K}$. The nematic-isotropic transition temperature
for 5CB is $T_{\mathrm{NI}} =308.5 \,\mathrm{K}$. We choose
$T=308.3\,\mathrm{K}$, corresponding to $Q_b \approx 0.31$, close
to the NI transition point, at which flocculation of colloidal
particles, driven by attractive interactions, normally occurs. The
nematic coherence length is $\xi = \left( 24 L_1 c/b^2
\right)^{1/2} \approx 10\, \rm nm $; it sets the smallest
length-scale in our system.

Three identical colloidal particles, each of which we take to be a
long cylinder of radius $R$ with the symmetry axis parallel to the
$z$ axis, are immersed into a nematic liquid crystal. The order
parameter at the system boundaries is fixed to the bulk order
parameter of the nematic phase; the director orientation at the
box boundaries is fixed along the $y$ axis. In this geometry the
colloidal particles are orthogonal to the director field far from
them (the other geometry is when the rods are aligned with the
nematic). The problem of equilibrium orientation is discussed in
Ref.~\cite{burylov:1994.a}.

The order parameter at the particle surfaces is fixed to the bulk
order parameter. We consider fixed homeotropic boundary conditions
at the particle surfaces, i.~e. the director orientation is fixed
along the surface normals.

Assuming translational symmetry of the system along the $z$ axis,
the equilibrium distribution of the order parameter tensor
$Q_{ij}$ minimizes the free energy functional $F = \int f dxdy$,
where  $f$ is the free energy density~(\ref{eq:nem_free_en}). The
functional $F$ is minimized numerically,  using finite elements
with adaptive meshing. The two-dimensional domain of integration
of the free energy density $f$ is triangulated and the functions
$Q_{ij}$ are linearly interpolated within each triangle. Then the
excess free energy $F_N$ is calculated as
\begin{equation}
F_N = L_z \int(f - f_b)dxdy,
\label{eq:el_nem_free_en}
\end{equation}
where $f_b = f(Q_b)$ and $L_z$ is the length of colloidal
particles. Here we will only talk about effective interactions per
length, $\bar{F}_N = F_N/L_z$, $\bar{F}^{(n)} = {F}^{(n)}/L_z$. It
will be understood that such scaling has been carried out, and we
shall omit the overbars in the text below.

\section{Results}
\subsection{Collinear geometry}

\begin{figure}
\includegraphics[width=4cm]{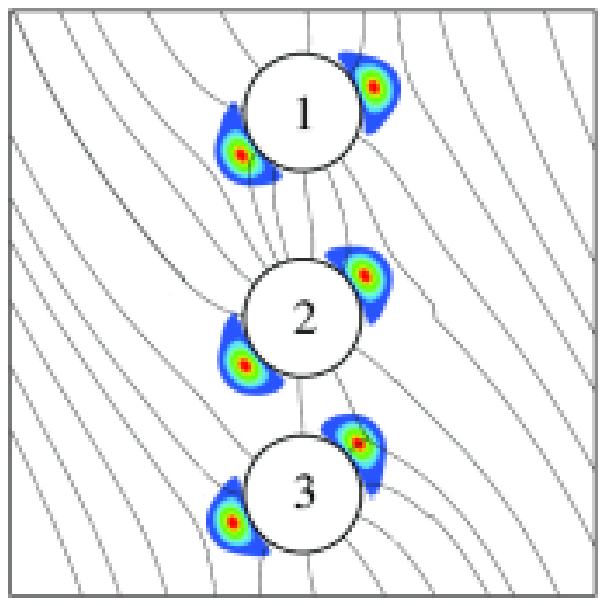}
\includegraphics[width=4cm]{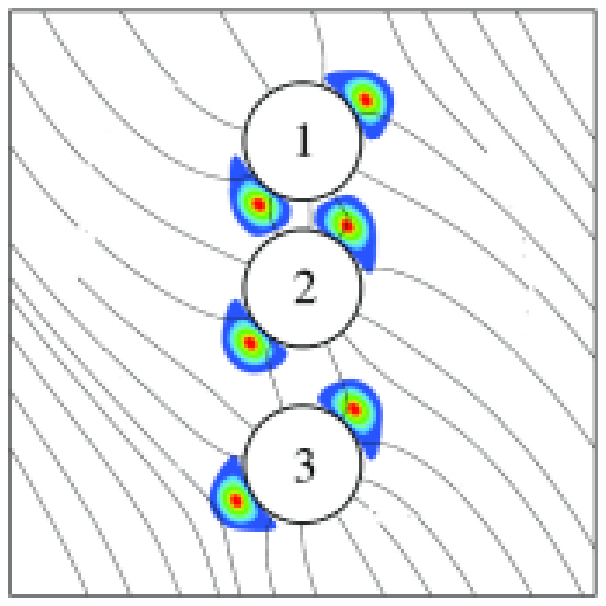}
\caption{ (Color on-line) Order parameter and director
distribution around colloidal particles with $R/\xi=10$. Left:
$r_{23} = 3R$, $r_{12} = 3.5R$; right: $r_{23} = 3R$, $r_{12} =
2.5R$. Director orientation far from the particles is parallel to
the line connecting particle centers. Note that only a small part
next to the particles is shown, the actual box size is $40 R
\times 40 R$.} \label{fig:snapshot_collinear}
\end{figure}
\begin{figure}
\includegraphics[width=8cm]{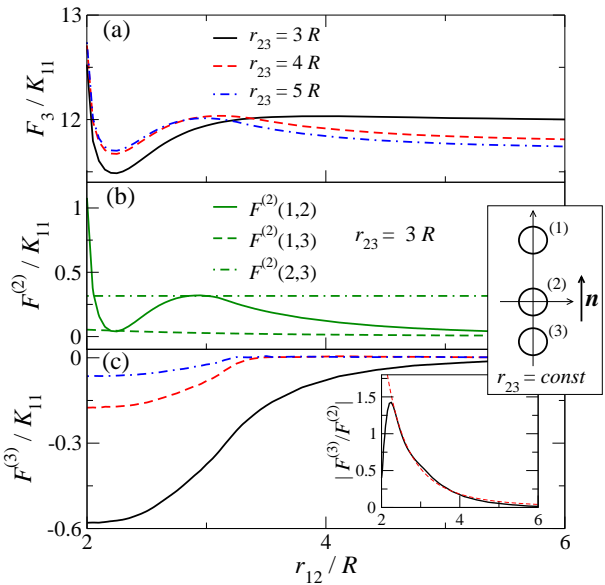}
\caption{ (Color on-line) Collinear geometry. Three colloids are
positioned on the line which is parallel to the director $\bm n$.
Particle radius $R/\xi = 10$. (a) Excess free energy $F_3$ as a
function of the separation between the particles (1) and (2). The
position of the particle (3) is fixed with respect to the particle
(2) at $3R$, $4R$, and $5R$. (b) Pair potentials for $r_{23}=3R$.
(c) Three-body contribution to $F_3$; solid, dashed and
dash-dotted lines correspond to $r_{23} = 3R, 4R$ and $5R$,
respectively. Inset shows the ratio between the three- and
two-body contributions and a power law fit of the decaying tail
for $r_{23}=3R$.} \label{fig:collinear}
\end{figure}

In this case the three colloids are positioned on a line parallel
to the nematic director $\bm n$, along the $y$ axis. Typical
director orientation and order parameter maps are shown in
Fig.~(\ref{fig:snapshot_collinear}). Each particle is accompanied
by two defect lines of strength $-1/2$, and the orientation of
each defect pair is tilted with respect to the (fixed) director
orientation far from the particles,
Fig.~(\ref{fig:snapshot_collinear}a). Note that if a single
colloidal particle is immersed in the nematic host, the
accompanying defects will be positioned on a line perpendicular to
the director orientation. The tilt angle (which breaks the mirror
symmetry of the system) is a pure two-body effect and is already
present in a system of two colloids~\cite{sivestre:2004.a}. At
small separations, the three particles in a collinear geometry can
serve as a model for a rod for which symmetry breaking has also
been observed~\cite{andrienko:2002.b}. When the distance between
the particles (1) and (2) is decreased (see
Fig.~(\ref{fig:snapshot_collinear}b)), the orientation of the
defects next to these particles tilts even more. However, this
does not affect the order parameter and the director distribution
around the third particle, i.~e. one might anticipate that in a
collinear geometry the contribution of the three-body potential to
the total interaction energy is rather small, in line with the
results of Guzm{\'a}n et.~al.~\cite{guzman04}.

To asses this contribution, the total interaction energy $F_3$ as
well as the pair $F^{(2)}$ and the triplet $F^{(3)}$ contributions
to it are shown in Fig.~(\ref{fig:collinear}). Here we fix the
distance between the particles (2) and (3) at $3R$, $4R$, and $5R$
and study the interaction of these two particles with the particle
(1). The dependence of $F_3$ on separation $r_{12}$ is similar for
all three curves: particle (1) is repelled from the particles (2)
and (3) at large distances and there is a barrier which does not
allow particle (1) to form a stable aggregate. The height of this
barrier decreases with the decrease of the distance between the
particles (2) and (3). Figure~(\ref{fig:collinear}b) shows that
the major contribution to the total interaction energy comes from
the pairwise interaction of particles (1) and (2), especially at
large separations of particles (2) and (3). However, for $r_{23} =
3R$ we have significant contribution of the three-body potential,
which is shown in Fig.~(\ref{fig:collinear}c). In fact, this
contribution is responsible for the decrease of the repulsive
barrier and will lead to more efficient clustering of the
colloids. For $r_{23}\lesssim 3R$ the pairs of defects around
particles (2) and (3) start to tilt. The tilt angle increases with
the decrease of $r_{23}$, affecting the effective interaction
between the pair and the third colloid~\cite{tasinkevych03}.

One can also see that in this geometry the three-body potentials
always provide an attractive contribution to the total interaction
energy. To emphasize the role of the three-body term, its ratio to
the sum of two-body potentials is shown in the inset of
Fig.~(\ref{fig:collinear}c) for  $r_{23}=3R$. It clearly shows
that at small separations, when rearrangement of the defects plays
a major role, the three-body term is dominating and therefore,
cannot be neglected. At larger distances the role of the
three-body term becomes less important decaying as a power law
$\mathcal{F}^{(3)}/\mathcal{F}^{(2)} \sim r_{12}^{-3.7}$.

\begin{figure}
\includegraphics[width=8cm]{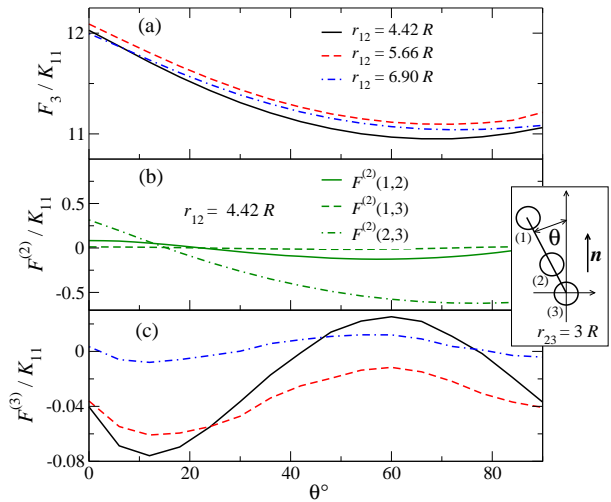}
\caption{ (Color on-line) Collinear geometry, dependence on the
tilt angle with respect to the nematic director: (a) excess free
energy $F_3$; (b) two-body potentials; (c) three-body potential,
solid, dashed and dash-dotted lines correspond to $r_{12} = 4.42R,
5.66R$ and $6.90R$, respectively. Particle radius $R/\xi = 10$.}
\label{fig:collinear_tilted}
\end{figure}

We also study how the two contributions change under the director
rotation. Figure~(\ref{fig:collinear_tilted}) shows the tilt angle
dependence of the excess free energy $F_3$, as well as two- and
three-body contributions to it. It is interesting that our
calculations predict an equilibrium configuration with a tilt
angle about $70^{\circ}$ with respect to the nematic director,
i.~e. there is also a torque acting on the line of colloids. The
tilt is due to the pairwise interactions: the three-body
contribution, in fact, has a \textit{maximum} at about
$60^{\circ}$ and favors almost collinear with the director
alignment (minimum at about $10^{\circ}$).

\subsection{Equilateral triangle geometry}

\begin{figure}
\includegraphics[width=8cm]{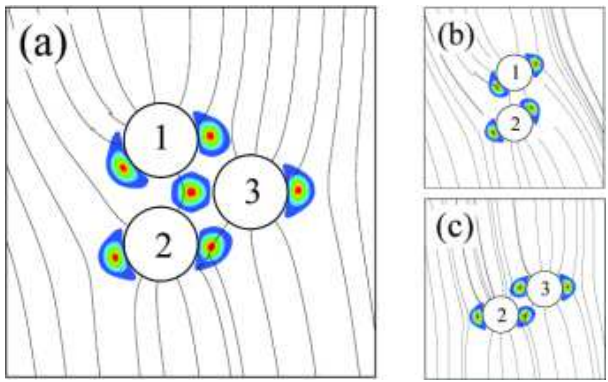}
\caption{ (Color on-line) Order parameter and director
distribution for the case of equilateral triangle geometry,
$r_{12}=r_{13}=r_{23}=2.8R$, $R/\xi = 10$. (a) three particles;
(b) the same geometry, but the particle (3) is removed; (c) the
particle (1) is removed. The director far from the particles is
parallel to the line connecting the centers of particles (1) and
(2). Only a small area next to the particles is shown.}
\label{fig:snapshot_equilateral}
\end{figure}

\begin{figure}
\includegraphics[width=8cm]{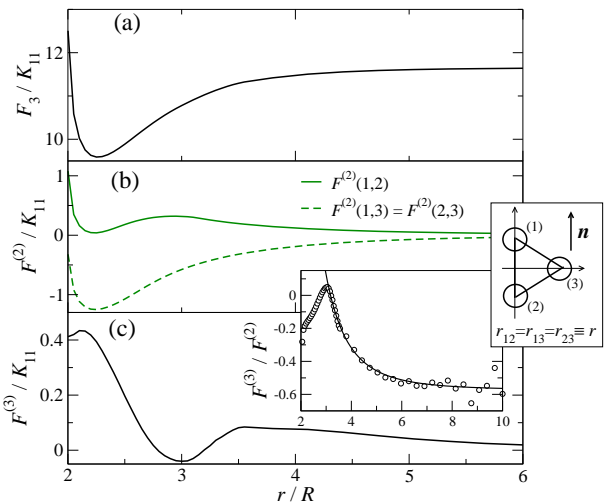}
\caption{ (Color on-line) Equilateral triangle geometry, $R/\xi =
10$. (a) total interaction energy $F_3$;  (b) two-body
contributions; (c) three-body contribution. Inset shows the ratio
between the three-and two-body contributions and a power law fit
($r^{-3.5}$) of the tail. } \label{fig:equilateral}
\end{figure}

The order parameter and the effective interaction potentials for
the equilateral triangle geometry are presented in
Fig.~(\ref{fig:snapshot_equilateral}) and
Fig.~(\ref{fig:equilateral}), respectively. Already the director
and order parameter profiles, depicted in
Fig.~(\ref{fig:snapshot_equilateral}), show that the three-body
potential is significant in this case: the position of the defects
and the director orientation  around three colloids (see
Fig.~(\ref{fig:snapshot_equilateral}a)) cannot be obtained as a
superposition of the corresponding two-particles profiles (see
Fig.~(\ref{fig:snapshot_equilateral}b) and
Fig.~(\ref{fig:snapshot_equilateral}c)), even at large
separations. Since in the equilateral triangle  geometry defect
structures are less separated than in the collinear one, a much
stronger three-body contribution is expected in this case.

In Fig.~(\ref{fig:equilateral}a) $F_3$ as a function of the
triangle size $r$ is shown, demonstrating attraction of the
particles at large $r$. Pairwise contributions show that the
particles (1) and (2) repel each other at large distances, while
pairs (2)-(3) and (1)-(3) attract.  The three-body contribution
is almost always positive, and has a minimum around $r=3R$.
Contrary to the collinear geometry, the ratio $F^{(3)}/F^{(2)}$,
which is shown in the inset of Fig.~(\ref{fig:equilateral}), does
not vanish at large $r$. Moreover, it converges to a constant
(negative) value. This means that at large distances the
three-body potential decays with the same power law as the
two-body ones. Therefore, it cannot be neglected even at large
separations between the colloidal particles.

\begin{figure}
\includegraphics[width=8cm]{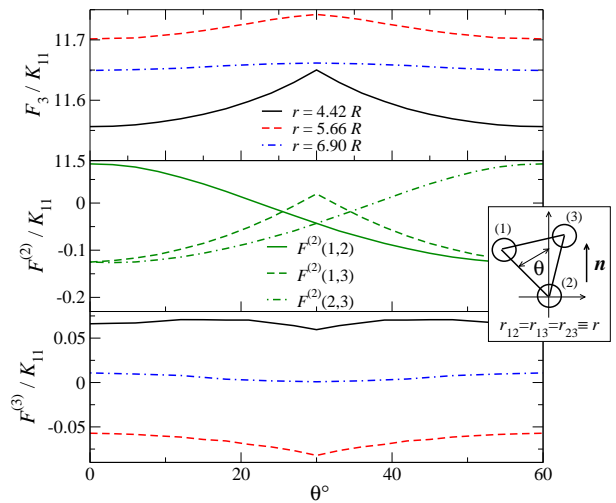}
\caption{ (Color on-line) Equilateral triangle geometry, tilt
angle dependence of the effective interaction energies. $R/\xi =
10$. Bottom panel: solid, dashed and dash-dotted lines correspond
to $r = 4.42R, 5.66R$ and $6.90R$, respectively.}
\label{fig:equilateral_tilted}
\end{figure}

The effective interaction potentials reveal rather weak dependence
on the tilt angle, which is shown in
Fig.~\ref{fig:equilateral_tilted}. The  peaks around $30^{\circ}$
are related to the defect rearrangement around the particles.

\subsection{Midplane geometry}

\begin{figure}
\includegraphics[width=8cm]{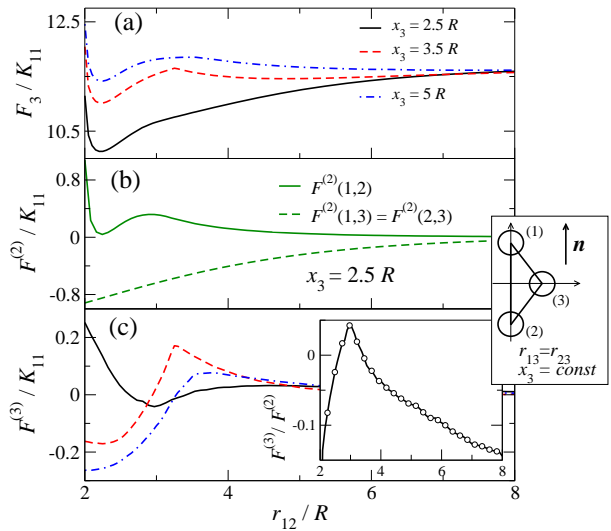}
\caption{ (Color on-line) Midplane geometry, $R/\xi = 10$: (a)
total interaction energy; (b) pairwise contributions; (c)
three-body contribution; solid, dashed and dash-dotted lines
correspond to $x_3 = 2.5R, 3.5R$ and $5R$, respectively. Inset
shows the ratio between the three-and two-body contributions for
$x_3 = 2.5R$.} \label{fig:midplane}
\end{figure}

In this  geometry the position of the particle (3) is fixed at a
distance $x_3$ from the line connecting particles (1) and (2) (the
geometry is depicted in Fig.~(\ref{fig:midplane})). The addition
of the third particle changes the repulsive interaction between
particles (1) and (2) to the attractive one, favoring their
flocculation. The $F^{(3)}/F^{(2)}$ ratio shows the same tendency
as in the equilateral triangle case: it does not reach zero at
large $r_{12}$, i.~e. the three-body potential is long-ranged and
provides about 10\% correction to the sum of pair potentials at
all distances.

\subsection{Four-body interactions}

\begin{figure}
\includegraphics[width=8cm]{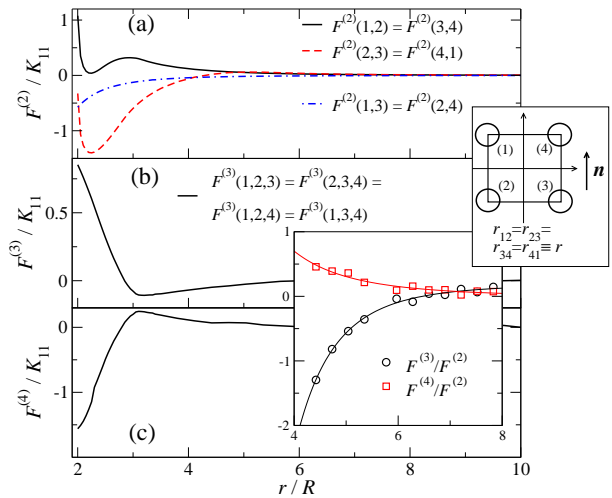}
\caption{ (Color on-line) Effective interaction potentials for
four colloidal particles in a square geometry, with $R/\xi = 10$.
(a) pairwise contributions; (b) three-body contributions; (c)
four-body contribution. Inset shows the ratio between the three-,
four- and two-body contributions with the power-law fits.}
\label{fig:square}
\end{figure}

To get the feeling of the next (four-body) term in the
expansion~(\ref{eq:el_free_en_decompos}) we considered a system of
four colloidal particles positioned at the vertices of a square.
Two-, three-, and four-body potentials are shown in
Fig.~(\ref{fig:square}). The inset shows the ratio of the total
three- and four-body contributions to the two-body one. The total
two- and three-body contributions to $F_4$ are calculated as
follows
\begin{eqnarray}
\nonumber
{F}^{(2)} &=& 2\left[
{F}^{(2)}(1,2) +
{F}^{(2)}(2,3) +
{F}^{(2)}(1,3)
\right], \\
\nonumber
{F}^{(3)} &=& 4{F}^{(3)}(1,2,3).
\end{eqnarray}

Fitting the ratios ${F}^{(i)}/{F}^{(2)} $ ($i=3,4$) to the power
law $(a_i/x)^{\gamma_i} + b_i$ in the region $r \in [6R..10R] $
yields $\gamma_{3} = 5.5$, $b_{3} = 0.18$  and $\gamma_{4} = 3.6$,
$b_{4} = -0.01$. Again, at short distances the contribution of the
three-body potentials to $F_4$ is comparable (and at small
distances even bigger) to the contribution of the two-body
potentials. Moreover, the fit shows that both contributions have
the same range and that three-body potentials amount up to 10\% of
the two-body terms even at large separations. The four-body term
is slightly smaller than the three-body one. However, the slow
convergence of the expansion~(\ref{eq:el_free_en_decompos}) does
not allow to make a quantitative estimate of a many-body
contribution that can be neglected.

\section{Conclusions}
Our calculations indicate that a collinear array of colloids (in
our case modeled as infinitely long rods) is not as stable as the
equilateral triangle configuration. The most stable equilateral
triangle configuration is when one of the sides of the triangle is
parallel to the nematic director.

Russ at.~el.~\cite{russ02} have shown that the triplet interaction
is a function of the summed distance $r_{12}+r_{23}+r_{31}$
between the three particles only. However, this empirical
observation does not hold in our case. The possible reasons are:
anisotropic nature of the interactions and presence of topological
defects, both of which are absent in a system of charged colloids.

The main message of this work is that in nematic colloids
many-body effects cannot be neglected. At short distances
rearrangements of accompanying defects strongly contribute to the
three- and four- body terms in the free energy expansion. In some
geometries (e.~g. equilateral triangle) the exponents of the power
law decay of the two- and three- body contributions with
interparticle distance are the same, i.~e. the higher order
contributions cannot be neglected even at large separations.

\acknowledgments Comments, remarks, and discussions with Markus
Deserno are greatly acknowledged.

\bibliography{paper}

\end{document}